# Raising and lowering operators for angular momentum quantum numbers $l$ in spherical harmonics

Q. H. Liu[*], D. M. Xun, and L. Shan

Key Laboratory for Micro-Nano Optoelectronic Devices of Ministry of Education, and State Key Laboratory for Chemo/Biosensing and Chemometrics,
and School for Theoretical Physics, and Department of Applied Physics, and Hunan University, Changsha, 410082, China.

**Abstract**

Two vector operators aimed at shifting angular momentum quantum number $l$ in spherical harmonics $|lm\rangle$, primarily proposed by Prof. X. L. Ka in 2001, are further studied. For a given magnetic quantum number $m$, specific states $|lm\rangle$ in spherical harmonics with the lowest angular momentum quantum numbers $l$ are obtained and the state with minimum angular momentum quantum number in whole set of the spherical harmonics is $|0,0\rangle$. How to use these states to generate whole set of spherical harmonics is illustrated.

Key Words: angular momentum, raising and lowering operators, quantum numbers

PACS numbers:   03.65.Ca

---

[*] School for Theoretical Physics, and Department of Applied Physics, Hunan University, Changsha, 410082, China; e-mail: quanhuiliu@gmail.com.

# I. INTRODUCTION

Testing an idea against a vital theoretical model usually enriches the understanding of both the idea and the model. In quantum mechanics, the ladder operator technique is widely used. For instance, the action of the angular momentum ladder operator $L_+$ and $L_-$ with definition $L_\pm \equiv L_x \pm iL_y$ on spherical harmonics $|lm\rangle$ raises and lowers respectively the magnetic quantum number $m$ by one while leaving the angular momentum quantum number $l$ unaltered. Then is there any ladder operator that shifts the values of $l$ in the spherical harmonics $|lm\rangle$?

Looking into literature, we can find that there are indeed results relevant to the solution to this problem. In 1980, Szpikowski and Góźdź pointed out in passing in the appendix A of their paper (Szpikowski and Góźdź 1980) an operator $O$ in the interacting boson nuclear model can diminish both $l$ and $m$ in $|lm\rangle$ with $m=l$ as $|l,l\rangle$ to another one $|l',l'\rangle$, where the operator $O$ is a polynomial of terms containing powers of $L_+$, $L_-$ and the tensor operators $T^{(k)}$ where superscript $k$ denotes the rank under rotational transformations. In 1994, with help of the tensor operators $T$, Shanker (Shanker 1994) showed that the raising and lowering operator $A_\pm = A_x \pm iA_y$ constructed form the Lenz vector operator $\mathbf{A} = (\mathbf{p}\times\mathbf{L} - \mathbf{L}\times\mathbf{p})/2 - \mathbf{r}$ acting on spherical hydrogen atom eigenstates $|nlm\rangle$ happens to be $A_\pm |nll\rangle = D_{ll}^\pm |n;(l\pm1);(l\pm1)\rangle$, where $D_{ll}^\pm$ are constants depending on $l$. Burkardt and Leventhal in 2004 demonstrated that the same relation $A_\pm |nll\rangle = D_{ll}^\pm |n;(l\pm1);(l\pm1)\rangle$ can be obtained without resorting to the tensor operator (Burkardt and Leventhal 2004).

As far as our knowledge goes, the first attempt to give a direct answer to the problem is due to Prof. Ka in 2001, who presented a derivation and observed that once acting on the spherical harmonics $|lm\rangle$ two vector operators (Ka 2001),

$$\mathbf{R}(l) = \frac{i}{\hbar}\mathbf{N}\times\mathbf{L} + (l+1)\mathbf{N}, \qquad \mathbf{Q}(l) = \frac{i}{\hbar}\mathbf{N}\times\mathbf{L} - l\mathbf{N}, \qquad (1)$$

are good enough to meet our need, where $\mathbf{N} \equiv \mathbf{R}/R$ is the direction operator for position. Explicitly, with defining $R_\pm(l) \equiv R_x(l) \pm iR_y(l)$, and $Q_\pm(l) \equiv Q_x(l) \pm iQ_y(l)$, Ka showed that

$$R_\pm(l)|lm\rangle = \sqrt{\frac{2l+1}{2l+3}}a(l+2,\pm m)|l+1,m\pm 1\rangle, \quad Q_\pm(l)|lm\rangle = \sqrt{\frac{2l+1}{2l-1}}a(l,\mp m)|l-1,m\pm 1\rangle, \qquad (2)$$

$$R_z(l)|lm\rangle = \sqrt{\frac{2l+1}{2l+3}}b(l+1,m)|l+1,m\rangle, \quad Q_z(l)|lm\rangle = -\sqrt{\frac{2l+1}{2l-1}}b(l,m)|l-1,m\rangle, \qquad (3)$$

where

$$a(l,m) = \mp\sqrt{(l+m)(l+m-1)}, \quad b(l,m) = \sqrt{(l+m)(l-m)}. \qquad (4)$$

Unfortunately, these two vector operators (1) are not full ones because they contain information of the state acted, i.e., the angular momentum quantum number $l$. Prof. Ka left the operator (1) almost finished. In fact, they become full operators once the following replacement is made in operators (1),

$$l\hbar \to \frac{\sqrt{4L^2/\hbar^2+1}-1}{2}\hbar = \frac{\sqrt{4\mathit{Ł}^2+1}-1}{2}\hbar, \qquad (5)$$

where $\mathit{Ł} \equiv \mathbf{L}/\hbar$, $\mathit{Ł}^2 \equiv L^2/\hbar^2$ are respectively dimensionless angular momentum and its square. Explicitly, the vector operators we deal with in this paper take following simple forms,

$$\mathbf{R} = i\mathbf{N}\times\mathit{Ł} + \mathbf{N}\frac{\sqrt{4\mathit{Ł}^2+1}+1}{2}, \quad \mathbf{Q} = i\mathbf{N}\times\mathit{Ł} - \mathbf{N}\frac{\sqrt{4\mathit{Ł}^2+1}-1}{2}. \qquad (6)$$

These two vector operators have not been reported before, and many properties are possibly unknown to us yet. However, the present paper is mainly concerned with an illustration that they are really raising and lowering operators under looking for.

This paper is organized as following. In section II, the fundamental commutation relations are presented, and in Section III, the quantum states of lowest quantum numbers are determined. In last section IV in addition to discussions we will briefly mention some interesting topics of further studies.

## II. FUNDAMANTAL COMMUTATION RELATIONS

It is accustomed (Hall and Mitchell 2002) to assume that operator $L \equiv \sqrt{L^2}$ is hermitian and satisfies

$$[L^2, L] = 0, \quad [L, \mathbf{L}] = 0. \tag{7}$$

Choosing the simultaneous eigenvalues of two operators $L^2, L_z$, we have,

$$L|lm\rangle = \sqrt{l(l+1)}\hbar|lm\rangle. \tag{8}$$

Note that a vector operator $\mathbf{V}$ by definition satisfies the commutation relations, (Sukurai 1994)

$$[V_i, L_j] = i\hbar \varepsilon_{ijk} V_k. \tag{9}$$

Moreover if $\mathbf{V} \cdot \mathbf{L} = 0$, it can be easily proved (Ka 2001),

$$(\mathbf{V} \times \mathbf{L}) \times \mathbf{L} = i\hbar(\mathbf{V} \times \mathbf{L}) - \mathbf{V}L^2. \tag{10}$$

As a consequence of these relations (9)-(10), we have

$$\left[L^2, \mathbf{N} \times \mathbf{L}\right] = 2i\hbar(\mathbf{N} \times \mathbf{L}) \times \mathbf{L} + 2\hbar^2 \mathbf{N} \times \mathbf{L} = -2i\hbar \mathbf{N}L^2. \tag{11}$$

Using relations (9) and (11), we have immediately following commutation relation,

$$\left[L^2, \mathbf{R}\right] = \hbar^2 \mathbf{R}\left(\sqrt{4L^2 + 1} + 1\right). \tag{12}$$

Acting of this relation (12) on both sides with spherical harmonics $|lm\rangle$, we find that operator $\mathbf{R}$ really shifts the angular momentum quantum number from $l$ to $l+1$,

$$L^2 \mathbf{R}|lm\rangle = \mathbf{R} L^2 |lm\rangle + \hbar^2 \mathbf{R}\left(\sqrt{4L^2 + 1} + 1\right)|lm\rangle = (l+1)(l+2)\hbar^2 \mathbf{R}|lm\rangle. \tag{13}$$

In other words,

$$\mathbf{R}|lm\rangle \propto |l+1, m'\rangle, \tag{14}$$

where magnetic quantum number $m'$ may differ from the original one $m$. Similarly, we have for operator $\mathbf{Q}$,

$$\left[L^2, \mathbf{Q}\right] = -\hbar^2 \mathbf{Q}\left(\sqrt{4L^2 + 1} - 1\right), \tag{15}$$

and it shifts the angular quantum number from $l$ to $l-1$,

$$\mathbf{Q}|l, m\rangle \propto |l-1, m'\rangle. \tag{16}$$

Next, in order to examine how magnetic quantum number changes on the action of operators $\mathbf{R}$ and $\mathbf{Q}$, we need to calculate commutation relations such as $[L_z, \mathbf{R}]$ and $[L_z, \mathbf{Q}]$. By utilization of the definition of the vector operator (9), we have

$$[L_z, R_z] = 0, \quad \text{and} \quad [L_z, Q_z] = 0, \tag{17}$$

and

$$[L_z, R_\pm] = \pm \hbar R_\pm, \quad \text{and} \quad [L_z, Q_\pm] = \pm \hbar Q_\pm. \tag{18}$$

Eqs. (14) and (17) show that $R_z$ and $Q_z$ are operators respectively raise and lower the quantum number $l$ in spherical harmonics $|lm\rangle$ by one while keeping the magnetic quantum number $m$ unchanged. Eqs. (14), (16) and (18) show that $R_\pm$ and $Q_\pm$ are operators respectively raise and lower $l$ in $|lm\rangle$ by one and also move the magnetic quantum number $m$ by $\pm 1$ respectively. Explicitly, we have after some calculations,

$$R_\pm |lm\rangle = \sqrt{\frac{2l+1}{2l+3}} a(l+2,\pm m) |l+1, m\pm 1\rangle, \quad Q_\pm |lm\rangle = \sqrt{\frac{2l+1}{2l-1}} a(l,\mp m) |l-1, m\pm 1\rangle, \tag{19}$$

$$R_z |lm\rangle = \sqrt{\frac{2l+1}{2l+3}} b(l+1,m) |l+1, m\rangle, \quad Q_z |lm\rangle = -\sqrt{\frac{2l+1}{2l-1}} b(l,m) |l-1, m\rangle. \tag{20}$$

Note that the operators used here heave nothing to do with the state acted whereas those used in (2)-(3) are only half-finished.

From Eqs. (19)-(20), we see that two pairs of operators $R_\pm$ and $L_\pm R_z$ are respectively equivalent, and so are two other pairs of operators $Q_\pm$ and $L_\pm Q_z$,

$$R_\pm \sim L_\pm R_z, \quad Q_\pm \sim L_\pm Q_z. \tag{21}$$

## III. DETERMINATION OF QUANTUM STATES WITH LOWEST QUANTUM NUMBERS

From Eqs (19)-(20), operators that can lower the angular momentum quantum number are $Q_j$, $(j=+,-,z)$. Hence there must exist kets satisfying,

$$Q_j |lm\rangle_j = 0, \quad (j=+,-,z). \tag{22}$$

In these cases, subsequent applications of the lowering operators will produce zero kets. In contrast, by acting on these kets with appropriate raising and lowering operators and multiplying by suitable normalization factors, we can produce an infinite even whole set of the kets. For instance, once we know a state $|lm\rangle_j = |lm\rangle$, another meaningful ket $|l+p, m-q\rangle$ with $p>0$, $q>0$ and $p\geq q$ can be gotten via firstly applying $(R_-)^q$ and secondly $(R_z)^{p-q}$ on the state $|lm\rangle$, i.e., $(R_z)^{p-q}(R_-)^q |lm\rangle \propto (R_z)^{p-q}|l+q, m-q\rangle \propto |l+p, m-q\rangle$. In the following, we need to solve equations in (22) and discuss the relations between solutions. We will deal with this problem in spherical polar coordinates $(\theta,\varphi)$ where $Y_{lm}(\theta,\varphi) = \langle \theta\varphi | lm\rangle$. Because the vector operator **R** or **Q** contain two variables $(\theta,\varphi)$, solutions to Eqs. (22) must depend on two quantum numbers. In general, we look for solutions that are simultaneously eigenvalue of $L_z$.

Firstly, we solve the differential equation $Q_z |lm\rangle_z = 0$ where the operator $Q_z$ takes the following form,

$$Q_z = \sin\theta \frac{\partial}{\partial \theta} - \cos\theta \frac{\sqrt{4Ł^2+1}-1}{2}, \tag{23}$$

and

$$Ł^2 = -\left( \frac{1}{\sin\theta} \frac{\partial}{\partial \theta} \sin\theta \frac{\partial}{\partial \theta} + \frac{1}{\sin^2\theta} \frac{\partial^2}{\partial \varphi^2} \right). \tag{24}$$

In spherical polar coordinates, $Q_z |lm\rangle_z = 0$ assumes the form with $\psi_z(\theta,\varphi) = \langle \theta\varphi | lm\rangle_z$,

$$\left( \frac{\sin\theta}{\cos\theta} \frac{\partial}{\partial \theta} + \frac{1}{2} \right) \psi_z(\theta,\varphi) = \frac{\sqrt{4Ł^2+1}}{2} \psi_z(\theta,\varphi), \tag{25}$$

or,

$$\left(\frac{\sin\theta}{\cos\theta}\frac{\partial}{\partial\theta}+\frac{1}{2}\right)\left(\frac{\sin\theta}{\cos\theta}\frac{\partial}{\partial\theta}+\frac{1}{2}\right)\psi_z(\theta,\varphi)=\left(\frac{L^2}{\hbar^2}+\frac{1}{4}\right)\psi_z(\theta,\varphi). \tag{26}$$

By standard method of separation of variables and with usual requirement of single-valuedness of the state function, general solution to Eq.(26) is given by,

$$\psi_z(\theta,\varphi)=\left(c_{1z}(\sin\theta)^m+c_{2z}(\sin\theta)^{-m}\right)e^{im\varphi}, \quad (m=0,\pm 1,\pm 2,....), \tag{27}$$

hereafter $c_i$ ($i=1z,2z,1+,2+,1-,2-.$) denote integration constants. The square integrability requires that the exponent of sine function can not be negative. The final result is simply with a normalization factor $c_z$

$$\psi_z(\theta,\varphi)=c_z(\sin\theta)^{|m|}e^{im\varphi}. \tag{28}$$

In terms of kets, $|lm\rangle_z=c_z\||m|,m\rangle$. Further decrease of the quantum number $l$ leads to $\||m|-1,m\rangle$ that can be identified as a zero ket. Therefore the state $\||m|,m\rangle$ really has the lowest angular momentum quantum number for the given magnetic quantum number $m$, implying that angular momentum quantum number $l\geq|m|$, or in other words for a given $l$, $m=-l,-l+1,...,l$.

Secondly, we deal with other two equations $Q_-|lm\rangle_-=0$ and $Q_+|lm\rangle_+=0$ in (22). Recalling the relation $Q_\pm \sim L_\pm Q_z$ in (21) i.e., $L_\pm Q_z|lm\rangle_z \sim Q_\pm|lm\rangle_z=0$, we find that one of the two independent solutions to each of these two equations is known. In fact, when $m\leq 0$, $Q_-|lm\rangle_z\,(=\||m|-1,m-1\rangle)$ is actually a zero ket, while $m>0$, $Q_+|lm\rangle_z\,(=\||m|-1,m+1\rangle)$ is also the zero ket. One can then easily verify that another independent solution to the differential equation $Q_-|lm\rangle_-=0$ is $\||m|+1,-|m|\rangle$, whereas it is $\||m|+1,|m|\rangle$ for $Q_+|lm\rangle_+=0$. So, the final solutions are,

$$|lm\rangle_\pm = c_{1\pm}\||m|,\pm|m|\rangle + c_{2\pm}\||m|+1,\pm|m|\rangle, \tag{29}$$

where upper and lower signs correspond to $Q_+|lm\rangle_+=0$ and $Q_-|lm\rangle_-=0$, respectively.

At first glance, three sets of solutions with their lowest angular momentum quantum numbers corresponding to three equations (22) are completely determined respectively. Among six solutions, only four as $\||m|,\pm|m|\rangle$ and $\||m|+1,\pm|m|\rangle$ are independent. We comment on the solution $\||m|+1,\pm|m|\rangle$ in (29) that is not the state of lowest angular momentum quantum number because it can be further lowered as $Q_z\||m|+1,\pm|m|\rangle \propto \||m|,\pm|m|\rangle$. Since the minimum of magnitude of the magnetic quantum number is zero, the minimum of the angular momentum quantum number is also zero. So, the state with minimum quantum number in whole set of spherical harmonics is $|0,0\rangle$. From this state $|0,0\rangle$, we can create any one $|l,m\rangle$ ($m=-l,-l+1,...,l$). Aside from coefficients the state $|l,|m|\rangle$ can be created via $(R_z)^{l-|m|}(R_+)^{|m|}|0,0\rangle \propto (R_z)^{l-|m|}\||m|,|m|\rangle \propto |l,|m|\rangle$; and the state $|l,-|m|\rangle$ can be created via $(R_z)^{l-|m|}(Q_+)^{|m|}|0,0\rangle \propto (R_z)^{l-|m|}\||m|,-|m|\rangle \propto |l,-|m|\rangle$.

## IV. CONCLUSIONS AND REMARKS

Developing one step further of the work of Prof. Ka (Ka 2001), we present the full form of the raising and lowering vector operators **R** and **Q** that shift the angular momentum quantum number in spherical harmonics $|lm\rangle$. Apparently, three lowering operators give six different states each of them with its lowest angular momentum quantum number, and only four of them are independent. Careful analysis shows that only two of them, $\||m|, \pm|m|\rangle$, bear the lowest angular momentum quantum numbers for a given the magnetic quantum number of magnitude $|m|$, and the state with minimum angular momentum quantum number in whole set of spherical harmonics turns out to be $|0,0\rangle$. Starting from this state $|0,0\rangle$, we can generate the whole set of the spherical harmonics with appropriate action of the raising and lowering operators.

The new operators (6) introduced in present paper may have wider and deeper respects worthy of future explorations. Among of them we mention a connection between the operators (6) and the coherent states defined on the sphere. As reviewed by Hall and Mitchell (Hall and Mitchell 2002), there are different forms of coherent states proposed by substantially different points of view. Among these coherent states that are defined by the eigenfunctions of annihilation or lowering operators, one is introduced by Kowalski and Rembielinski who presented the normalized vector operators (Kowalski and Rembielinski, 2000),

$$\mathbf{Z} \equiv i\sqrt{e}\,\frac{2\sinh\left(\sqrt{4\mathit{Ł}^2+1}/2\right)}{\sqrt{4\mathit{Ł}^2+1}}\mathbf{N}\times\mathbf{Ł} + \sqrt{e}\left(\cosh\left(\sqrt{4\mathit{Ł}^2+1}/2\right) - \frac{\sinh\left(\sqrt{4\mathit{Ł}^2+1}/2\right)}{\sqrt{4\mathit{Ł}^2+1}}\right)\mathbf{N}. \qquad (30)$$

This operator is identical to neither $\mathbf{R}$ nor $\mathbf{Q}$ but their combination with some coefficients depending on operator $\sqrt{4\mathit{Ł}^2+1}$. For instance one can prove that

$$Z_z \equiv R_z \exp\left(-\frac{\sqrt{4\mathit{Ł}^2+1}+1}{2}\right)\frac{1}{\sqrt{4\mathit{Ł}^2+1}} + Q_z \exp\left(\frac{\sqrt{4\mathit{Ł}^2+1}-1}{2}\right)\frac{1}{\sqrt{4\mathit{Ł}^2+1}}. \qquad (31)$$

What we ascertain now is that operators $\mathbf{Z}$, $\mathbf{R}$ and $\mathbf{Q}$ are formed by combination of $\mathbf{N}\times\mathbf{Ł}$ (or $\mathbf{Ł}\times\mathbf{N}$) and $\mathbf{N}$ with some coefficients depending on operator $\sqrt{4\mathit{Ł}^2+1}$, and there is no simple linear relation in between. Furthermore, there are certainly relations between the operators (6) and the tensor operators $T^{(k)}$ or the Lenz vector operators for they also play roles in shift the angular momentum quantum number. All these issues will be discussed in detail in near future.

**ACKNOWLEDGMENTS**

This subject is supported by "973" National Key Basic Research Program of China (Grant No. 2007CB310500). The first author is grateful to Prof. X. L. Ka, Beijing Normal University, for help discussions.